\documentclass[referee]{aa} 
\begin{document}

\title{The Impact of Gravitational Redshift on Systemic Water Maser Emission in NGC 4258}

\author{Marco Spaans\inst{1}}

\institute{$^1$Kapteyn Astronomical Institute, P.O. Box 800, 9700 AV
Groningen, The Netherlands; spaans@astro.rug.nl}

\date{Received; accepted}

\abstract{
NGC 4258 is well known for its 22 GHz water maser features that appear
to reside in a thin Keplerian disk orbiting at $R_0=0.13$ pc.
The influence of gravitational redshift on
the level of maser saturation and the line strength and width
of {\it systemic} water maser emission is studied.
The effective gravitational velocity shift
{\it along the gain path through the disk} amounts
to about twice the thermal velocity width of H$_2$O
molecules at $500$ K for a masing ring of thickness $0.2R_0$.
This GR induced frequency shift has a sign {\it opposite} to the shift caused
by the projected velocity difference vector of Keplerian motion on the
{\it blue} side of the systemic maser emission and has the {\it same}
sign on the {\it red} systemic side.
Consequently, the negative optical depth that a line photon sees as it
traverses the masing disk along a path of minimal Keplerian velocity gradient
is larger on the blue side.
Note that one is {\it not} concerned here with an {\it overall} shift in the
maser spectrum caused by the fact that the masing ring as a whole is located
in the black hole's potential well, i.e., it is the {\it change} in the
gravitational potential along the gain path that matters.

It is found, for a Schwarzschild metric, that
saturation is enhanced (suppressed) for systemic
blue-shifted (red-shifted) maser features by gravitational redshift of line
photons. This asymmetry should be observable for NGC 4258. That is,
approaching peak flux systemic features are, {\it on average}, i.e., over a
period of several years, brighter, due to saturation, and wider
($\sim 2$ km/s), due
to rebroadening, than their receding counterparts ($\sim 1$ km/s).
The effects are consistent with observations that span several years, i.e.,
over periods longer than needed for disk material to move through the systemic
emission region.
A completely definitive confirmation of this effect can follow by combining
all available systemic 22 GHz data over the past two decades.
\hfill\break
\noindent
{\bf keywords.} galaxies: individual: NGC 4258 -- ISM: molecules -- molecular processes -- radiative transfer

}

\titlerunning{Water Masers in NGC 4258 - GR and Saturation}
\maketitle

\section{Introduction}

Bright 22 GHz water masers have been detected toward NGC 4258 (M106) and other
galaxies (Nakai et al.\ 1993). For the AGN NGC 4258 the detection of both
high-velocity, $\sim -400$ and $\sim 1400$ km/s, and systemic, $\sim 500$
km/s, emission (LSR) have lead to the
picture of a (warped by $6^o$) Keplerian accretion disk that rotates, at a
distance of about 0.1 pc, with
$\sim 900$ km/s around a massive ($M\approx 3.5\times 10^7M_\odot$) black
hole (Moran et al.\ 1995).
The overall appearance of the spectra is spiky and individual maser features
have FWHM widths of $1-2$ km/s (Greenhill et al.\ 1995, 1996).
The excitation (pumping) of the masers involves the
the efficient formation of H$_2$O at temperatures in excess of a
few hundred Kelvin at H$_2$ densities $n>10^8$ cm$^{-3}$
(Neufeld et al.\ 1994). The appearance of high-velocity
and systemic maser emission, as well as the presence of absorption in the
central feature, has been explained by Watson \& Wallin (1994) and Wallin et
al.\ (1998) for unsaturated amplification. Here, the effects of saturation and
gravitational redshift caused by the central black hole are studied.

\section{Model Description}

A thin Keplerian accretion disk, viewed edge-on, with a height $H=10^{15}$ cm
and radial coordinate $r$, is adopted as in Wallin et al.\ (1998).
For the impact parameter $b$ one can write the optical
depth of unsaturated masing gas as
$$\tau_{\delta v}(b)=\alpha\int exp[-\delta v(b)^2/v^2_D] ds,\eqno(1)$$
where $\alpha$ is the (constant) maser opacity,
the frequency obeys $\nu =\nu_0(1-\delta v/c)$ for the rest frequency
$\nu_0$ and the equivalent velocity difference $\delta v$.
Gravitational redshift adds a contribution $\delta v_{GR}=c r_g (1/r-1/r')$
for the Schwarzschild radius $r_g=2M$ in geometric units and the black hole
mass $M$. Keplerian motion follows $b(GM)^{0.5}/r^{1.5}$.
Radial gas motion (inward drift $<0.007$ km/s)
is ignored in this work (Moran et al.\ 1995). One finds
$r_g=1.1\times 10^{13}$ cm. The Doppler width $v_D=(v^2_{th}+v^2_{tu})^{0.5}$
contains the thermal width $v_{th}=(2kT/m)^{0.5}=0.7$ km/s of the H$_2$O gas
molecules with mass $m$ at $T=500$ K as well as turbulent motions $v_{tu}$.
It is assumed that the systemic
masering water molecules are located at $R_0=0.13$ pc
over a length scale $\Delta r=0.2R_0$, while the entire masing disk runs to
$2R_0$ (Greenhill et al.\ 1995; Wallin et al.\ 1998, 1999).
The Keplerian rotational velocity ranges between
$v_{rot}\sim 800-1100$ km/s from the outside to the inside of the masing
disk (Hashick et al.\ 1994).
The unsaturated gain $\gamma$ at line center is taken as
$ln(\gamma )\sim 11$ at LSR velocities of $\pm 20$ km/s (with
respect to systemic), which yields $ln(\gamma )\sim 24$ at about $\pm 900$
km/s. These numbers roughly reproduce a typical observed flux density of
about 2 Jy
with a temperature for the (unresolved) background continuum
source of $T_c\sim 10^7$ K for $b\sim 0$ (Wallin et al.\ 1998) and
$T_c\sim 10$ K for $b\ge 1$ (almost self-amplification; Watson \&
Wallin 1994), where $b$ is normalized to $R_0$.
It is assumed that a layer of non-inverted H$_2$O molecules outside or inside
of the masing gas, but with the same Doppler width and a large ($>25$ at $b=0$)
optical depth, is reponsible for the absorption
around the systemic velocity (Wallin et al.\ 1998).
The distance to NGC 4258 is 6.4 Mpc (Miyoshi et al.\ 1995).
The measured systemic velocity of $472 \pm 4$ km/s
derives from H$\alpha$ observations (Cecil et al.\ 1992).
The results presented below do not depend strongly on the assumed model
parameters.

\section{Asymmetric Systemic Emission: Gravitational Redshift and Saturation}

The rate of de-excitation due to stimulated emission, $R_s$, is,
for an arbitrary point $q$,
$$R_s(q)=\int d\Omega R_s(q,\Omega )=\int d\nu c^2A/8\pi h\nu_0^3\int\phi (\nu ) I_{\nu}(\Omega )d\Omega\eqno(2)$$
and depends on the magnitude of the velocity gradient along the line of sight
since this determines which path segment contributes to the frequency and
solid angle integrals.
In this, $A$ is the Einstein A coefficient of the 22 GHz maser transition,
$\Omega$ is the solid angle and $\phi$ is the absorption profile centered at
$\nu_0$.

Saturation for the 22 GHz maser line
occurs when stimulated emission competes with the effective decay
rate, $\Gamma\sim 1$ s$^{-1}$ ($\sim 0.1$ s$^{-1}$ {\it only} if infrared
transitions are very optically thick in all directions),
of masing states due to all other processes.
The rate for velocity relaxation, $R_v$ (set by the lifetimes of infrared
decays), is at least as rapid and $R_v\sim 2$ s$^{-1}$ at $n=10^9$
cm$^{-3}$ and $500$ K (Anderson \& Watson 1993). Rebroadening of maser lines is
important when $R_s$ exceeds $R_v$ (Nedoluha \& Watson
1991). For NGC 4258, observations (including $R_0$ and $\Delta r$)
indicate that $R_s\sim\Gamma$ (Wallin et al.\ 1998) and, hence, a
significant increase in $R_s$ would strongly impact the saturation and line
shapes of maser features.

One obtains for the step function $\theta$ and $\nu_D=\nu_0v_D/c$ that
$$\phi (\nu )\approx [\theta (\nu_0-\nu_D )-\theta (\nu_0+\nu_D )]/2\nu_D\eqno(3)$$
and, to a good approximation,
$$R_s(q)=\int d\Omega R_s(q-L[g],\Omega )e^{-\alpha\int_{q-L[g]}^qdp}\eqno(4)$$
for the velocity gradient $g=dv/dp$ along a path parameterized through
$p(q)$ and
$L=v_D/|g|$ the local Sobolev length at line center.
If one dominant beam exists, the problem becomes effectively one-dimensional,
and one can concentrate on a single direction and work only with
$R_s(q,\Omega )$. When one applies the variational principle to find the
change in the Sobolev length that results from Keplerian motion and
gravitational redshift, the result is
$$\delta R_s(\Omega )\approx R_s(\Omega )e^{\tau_K\delta L /L}\eqno(5)$$
with $\delta L\equiv L[KGR]-L[K]$,
in which $K$ indicates Keplerian motion and $KGR$ combined Keplerian motion
and gravitational redshift.
Evaluation of $\delta L$ is straightforward through
$$\delta v_{GR}=c r_g(1/r-1/r')\ {\rm km/s},\eqno(6)$$
$$\delta v_K=(GM)^{0.5}b R_0(1/r^{1.5}-1/r'^{1.5})\ {\rm km/s},\eqno(7)$$
and indicates that gravitational redshift compensates the Doppler shift that
results from projected Keplerian motion on the approaching side of the
accretion disk. The converse is true for the receding side.
Hence, for the {\it systemic} maser features saturation is both enhanced and
suppressed. It can be concluded immediately that blue-shifted systemic maser
features should be, {\it on average}, brighter than their red-shifted
counterparts.
The effect of gravitational redshift is quite small for the high-velocity
features since only a narrow range in $|r-r'|$ is probed.
Also, special relativistic effects are neglected since {\it changes} in the
$\gamma$ factor are extremely small due to the required velocity coherence
over a line width of 1-2 km/s.

Over a scale of $\Delta r$ at $b=0$, gravitational redshift amounts to
$\delta v_{GR}=1.4$ km/s. It is further assumed that $v_{tu}=v_s$ with the
sound speed $v_s=1.3$ km/s at $500$ K.
A Monte Carlo method, see Spaans \& van Langevelde (1992) for details,
is used to evaluate $\delta R_s$.
It is found for the adopted parameters that the average Sobolev length $L$
increases for the approaching systemic features by about 30\%, over $\Delta r$.
This causes the decay rate due to
stimulated emission to increase from $R_s\sim 1$ s$^{-1}$ to
$R_s(-)\sim 9$ s$^{-1}$ for blue-shifted maser features over the observed LSR
velocity range, and with respect to the originally adopted gain $\gamma$.
Conversely, one still finds $R_s(+)\sim 0.8$ s$^{-1}$ for red-shifted maser
emission along the longest path, i.e., the path that originates at about a
velocity width from line center. Hence, since the adopted parameters for
$R_0$, $\Delta r$, $M$ and $\gamma$ are motivated by
observations, it can be concluded generally that blue systemic features should
be impacted more strongly by saturation than red systemic features.
Note in this that a saturated maser source becomes a ''linear convertor'' for
$R_s(-)\ge\Gamma /h$ with $h\sim 1/2$ and thus
that {\it the brightest} blue systemic features should have peak flux densities
that are larger than their red counterparts by a factor of
$F\sim\Gamma /(hR_s(+))\sim 2.5$, consistent with
observations of peak flux densities ($\sim 5$ Jy) of individual blue
systemic maser features.

For the blue-shifted systemic maser features in NGC 4258 one expects, in the
(almost) saturated limit, a typical FWHM line width of
$\Delta v\le 1.67v_D=2.5$ km/s,
but in any case, at least, $\Delta v\ge v_D=1.5$ km/s (Emmering \& Watson
1994; Nedoluha \& Watson 1991),
consistent with observations (Moran et al.\ 1995; Greenhill et al.\ 1995).
The FWHM width of a typical red-shifted systemic maser feature
decreases only modestly with $R_s$ and is $\sim 0.6v_D=1.0$ km/s.
Hence, the brightest red-shifted systemic features should be, {\it on average},
narrower by a factor of $\sim 2$ than the brightest blue-shifted systemic
features.

It is concluded that turbulence, which is likely
present in accretion disks (Afshordi et al.\ 2004), need not be very
large, i.e., be nearly subsonic, in order to explain the typical FWHM
widths, 1-2 km/s, of emission spikes
in observed spectra of NGC 4258. Supersonic turbulence decays rapidly, while
subsonic turbulence is easier to maintain. This is consistent with
observations that the disk is not much thicker than $\sim 10^{15}$ cm
(Moran et al.\ 1995) given that turbulence alone can support a disk with height
$H\sim v_{tu}R_0/v_{rot}=5\times 10^{14}$ cm.
Finally, this result differs somewhat from
that of Wallin et al.\ (1998), who argue for a larger turbulent
velocity in an unsaturated medium to reproduce the observed maser feature
FWHM widths, but agrees with the tilted disk model of Wallin et al.\ (1999).

\section{Observational Comparison and Caveats}

Observations indicate that
blue-shifted systemic emission features tend to be brighter and
wider, {\it on average}, than receding systemic features
(Hashick et al.\ 1994, their figure 2; Wallin et al.\ 1998, their figure 3).
It is stressed here that the data displayed in figure 2 of
Hashick et al.\ (1994) is {\it cumulative} over a period of 5 years, 1987-1991.
Over such a period any fluctuation in the ambient conditions caused, e.g., by
turbulence, would travel from about -25 to +25 km/s in the systemic emission
region, approximately 0.005 pc, {\it if} the fluctuation moves with the bulk
Keplerian motion.
That is, such a fluctuation would be amplified on both the blue {\it and} red
side of the systemic emission over a period of a few years.

Conversely, if amplified fluctuations do not follow Keplerian motion, then it
is remarkable that they are all located on the approaching side of the disk.
I.e., individual maser features on the blue side are brighter and wider,
$>50$\% in flux and about a factor of two in line width (Hashick et al.\ 1994),
than their red counterparts.
Also, fluctuations driven by turbulent motions have typical
wavelengths of the order of $H=10^{15}$ cm (Stone et al.\ 1996), while the
height of the disk is generally much smaller than the extent
$\Delta r=8\times 10^{16}$
cm over which maser amplification takes place. So a large number number of
fluctuations is expected to contribute to the systemic maser emission, on both
the blue and red side.
Hence, the asymmetry between red and blue systemic emission that is present in
the Hashick et al.\ {\it cumulative} signal is unlikely to be random.
In this work, these observations are explained by general relativity and
saturation. Clearly, it is necessary to extend the Hashick et al.\ procedure
for epochs upto the present. Note in this the respective integrals under
the blue and red parts of the of Hashick et al.\ systemic maser emission.
The surface area underneath the red systemic side
is somewhat larger than that of the blue side. This implies that the
{\it total} flux, i.e., bright features plus pedestal emission, on the blue
systemic is set by a pumping rate and/or background radiation source that is
not larger than that of the red side. Together with the much stronger and
wider (saturated) maser features on the blue side this points to
competitive gain among different maser beams on the blue systemic side
(Alcock \& Ross 1985).

The strongest assumption that has been made in this work is that of constant
excitation conditions. Still, models that invoke X-rays to heat the gas
and drive ion-molecule reactions that lead to (warm) water favor
oblique illumination, which yields constant excitation
condition when viewed edge-on (Neufeld et al.\ 1994).
Hence, excitation conditions may indeed be fairly constant.
Future work will concentrate on a detailed analysis of line profiles and
excitation conditions in the presence of saturation. In addition, the impact
of beam crossing (Spaans \& van Langevelde 1992) on the shape of the line
profile will be investigated then.

Accretion disks around more massive black holes are desirable to
further test general relativistic effects. Conversely, the effects mentioned
here should be (statistically) smaller in lower mass systems like NGC 1068
($M\sim 10^7M_\odot$). Also, water maser galaxies much farther away than NGC
4258 do not allow the motions of individual maser spots to be followed.
However, the observation of asymmetries in the systemic maser emission may then
still lead to a (model dependent) determination of the black hole mass.
Finally, for NGC 4258, gravitational redshift would not
lead to observational effects if $\Delta r<0.08R_0$.


\end{document}